\documentclass[a4paper,10pt,twoside,bibnote]{cpc-hepnp}

\usepackage{multicol}
\usepackage{graphicx}
\usepackage{booktabs}
\usepackage{amssymb,bm,mathrsfs,amscd}
\usepackage[tbtags]{amsmath}
\usepackage{lastpage}
\usepackage{multirow}
\usepackage{textcomp}
\usepackage{epstopdf}
\usepackage{float}
\usepackage{footnote}
\usepackage[colorlinks=true,citecolor=blue,linkcolor=blue,urlcolor=blue]{hyperref}

 \newcommand{\BESIIIorcid}[1]{\href{https://orcid.org/#1}{\hspace*{0.1em}\raisebox{-0.45ex}{\includegraphics[width=1em]{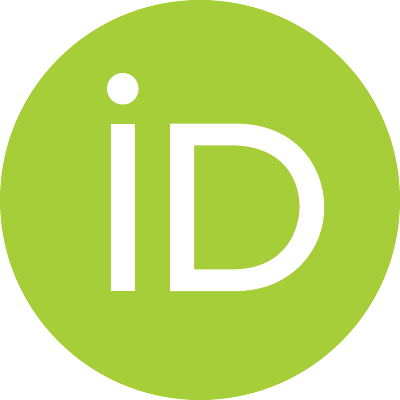}}}}

\usepackage{orcidlink}

\title{A High Performance Partial Wave Analysis Framework \\ for Hadron Spectroscopy}

\begin{document}

\setlength{\abovecaptionskip}{4pt plus1pt minus1pt}   
\setlength{\belowcaptionskip}{4pt plus1pt minus1pt}   
\setlength{\abovedisplayskip}{6pt plus1pt minus1pt}   
\setlength{\belowdisplayskip}{6pt plus1pt minus1pt}   
\addtolength{\thinmuskip}{-1mu}                       
\addtolength{\medmuskip}{-2mu}                        
\addtolength{\thickmuskip}{-2mu}                      
\setlength{\belowrulesep}{0pt}                        
\setlength{\aboverulesep}{0pt}                        
\setlength{\arraycolsep}{2pt}                         


\author{
Benhou Xiang$^{1,2}$\thanks{xiangbh@ihep.ac.cn}       
Shuangshi Fang$^{1,2,3}$\thanks{fangss@ihep.ac.cn}    
Beijiang Liu$^{1,2}$\thanks{liubj@ihep.ac.cn}         
}
\maketitle
\address{
$^{1}$Institute of High Energy Physics, Chinese Academy of Sciences, Beijing 100049, China\\
$^{2}$University of Chinese Academy of Sciences, Beijing 100049, China\\
$^{3}$School of Physics, Henan Normal University, Xinxiang 453007, China
}

\begin{abstract}
Partial wave analysis (PWA) is a key method in hadron physics for extracting the properties of hadronic resonances. As experimental statistics increase, conventional implementations face substantial challenges in memory usage and computational speed. We present \texttt{CTPWA}, a high-performance PWA framework based on the covariant tensor formalism. This framework incorporates extensive precomputation and caching mechanisms, together with fully GPU-based likelihood computation and minimization. These optimizations accelerate fitting by two orders of magnitude relative to autograd-based GPU PWA programs, making high-statistics partial wave analyses feasible in high-precision experiments such as BESIII.
\end{abstract}

\section{Introduction}

Hadron spectroscopy provides a critical window into the non-perturbative regime of Quantum Chromodynamics (QCD) through the study of bound states of quarks and gluons. In the conventional quark model, ordinary hadrons are color-singlet states composed of a quark--antiquark pair (mesons) or three quarks (baryons). QCD, however, also allows more complex configurations beyond this picture, including glueballs, hybrids, tetraquarks, and pentaquarks~\cite{ParticleDataGroup:2024cfk}. Mesonic exotic candidates are traditionally classified into two categories~\cite{Gross:2022qcd}: states with special quantum numbers, either in $J^{PC}$ or in flavor, that cannot be generated by an ordinary quark--antiquark pair; and crypto-exotic states, which share the quantum numbers of conventional mesons and therefore readily mix with them. Beyond the meson sector, analogous exotic signatures also appear in baryon spectra, notably in pentaquark candidates. Determining the genuine nature of these resonances across both meson and baryon systems, whether they are compact multiquark states, hadronic molecules, hybrids, or admixtures thereof, is therefore crucial for mapping the effective degrees of freedom of QCD in the low-energy domain and remains one of the foremost open questions in hadron spectroscopy.

Moreover, these resonances are generally too short-lived to be observed directly, and their properties must be inferred from the final-state particles in the processes in which they are produced. In multi-body decays, multiple intermediate states usually contribute to the same event sample and interfere with one another. Partial wave analysis (PWA) addresses this problem by modeling the decay amplitude and extracting the quantum numbers, masses, widths, and couplings of the contributing states through an unbinned maximum-likelihood fit.

The computational cost of PWA, however, becomes a serious limitation in the high-statistics era. The amplitude must be computed for every event, often for multiple helicity configurations and many interfering partial waves, while the event sample may easily reach $10^5$--$10^6$ events, although the number of free fit parameters is usually moderate. Because the events are mutually independent, this computation is well suited to parallel processing. GPU-based frameworks such as GPUPWA~\cite{Liu:2014qxa} and TF-PWA~\cite{Jiang:2024vbw} have therefore been developed to offload these calculations onto graphics processing units; these frameworks have been widely adopted in the community and underpin many analysis workflows at experiments.

Even with GPU offloading, PWA of large samples faces severe limitations. Device memory is one such limitation. Amplitude arrays or computational graphs are built and cached for all events, and the graph history, i.e. the intermediate values kept on the device for the backward pass, accumulates during the optimization. As a result, memory usage substantially exceeds that of the data themselves, and the device memory can be exhausted before the fit even starts. Speed is another limitation. The amplitude is recomputed at every iteration, so large amounts of data and intermediate results are continuously transferred between the CPU and the GPU, and the fit time is dominated by data-transfer and kernel-launch overheads. In addition, implementations built on PyTorch~\cite{pytorch} or TensorFlow~\cite{tensorflow2016} re-trace the computational graph at every iteration, an overhead that is unnecessary for a likelihood that is exactly quadratic in the couplings.

In this work, we present \texttt{CTPWA}, a lightweight PWA framework based on the covariant tensor amplitude formalism~\cite{Zou:2002ar,Jing:2023rwz,Jing:2024mag}. To address the difficulties described above, the framework adopts extensive precomputation and caching mechanisms, as well as fully GPU-based likelihood computation and minimization. The key observation underlying the method is that a substantial part of the decay amplitude is determined solely by the event kinematics and the decay topology and is therefore independent of the fit parameters. These kinematic building blocks can be precomputed once and reused throughout the fit, and only the resonance propagators and coupling combinations need to be updated during minimization. We exploit this observation fully by casting the likelihood into a compact matrix form whose gradient and Hessian follow analytically, thereby avoiding the overhead of graph-based automatic differentiation. The likelihood evaluation, its derivatives, and the limited-memory Broyden--Fletcher--Goldfarb--Shanno (L-BFGS) minimization~\cite{Liu:1989esw} run entirely on the GPU. As a result, the required device memory is fixed and known before the fit, no memory is allocated during the optimization, and traffic between the host and the device is reduced to the small parameter vectors (couplings and resonance parameters).

This paper is organized as follows. Section~\ref{ctamp} presents the likelihood construction. Section~\ref{method} describes the precomputation and caching mechanisms, as well as the GPU-based likelihood computation and minimization. Section~\ref{bench} reports the computational performance and a realistic fit of $J/\psi\to K^+K^-\eta$ with one million events. Section~\ref{summary} provides a summary.

\section{Likelihood construction}
\label{ctamp}

A partial wave analysis determines resonance parameters from the measured four-momenta of the final-state particles using an unbinned maximum-likelihood fit~\cite{Liu:2014qxa,Jiang:2024vbw}. Each event is characterized by the four-momenta $x \equiv \{p_1,\ldots,p_n\}$ of its final-state particles. The likelihood for the data sample is given by the product of the per-event probability densities,
\begin{equation}
\mathcal{L}(x;\boldsymbol{\theta}) = \prod_{i=1}^{N_{\rm data}} \mathcal{P}(x_i;\boldsymbol{\theta}),
\label{eq:lik}
\end{equation}
where the fit parameters $\boldsymbol{\theta}=(\boldsymbol{\theta}_a,\boldsymbol{\theta}_r)$ comprise the coupling parameters $\boldsymbol{\theta}_a$ for the decay chains and the resonance parameters $\boldsymbol{\theta}_r$. The probability density for an event is the decay intensity normalized by its phase-space integral.
\begin{equation}
\mathcal{P}(x;\boldsymbol{\theta}) = \frac{I(x;\boldsymbol{\theta})}{\sigma(\boldsymbol{\theta})}, \qquad
\sigma(\boldsymbol{\theta}) = \int I(x;\boldsymbol{\theta})\,\mathrm{d}\Phi,
\label{eq:prob}
\end{equation}
where $I$ is the event intensity, i.e., the squared modulus of the decay amplitude summed over all physical polarizations, and $\mathrm{d}\Phi$ is the phase-space volume element. The normalization integral $\sigma$ cannot be evaluated analytically and is therefore estimated using a phase-space Monte Carlo sample of $N_{\rm phsp}$ events.
\begin{equation}
\sigma(\boldsymbol{\theta}) \approx \frac{1}{N_{\rm phsp}} \sum_{j=1}^{N_{\rm phsp}} I_j(\boldsymbol{\theta}).
\label{eq:sigma}
\end{equation}
Substituting Eq.~(\ref{eq:prob}) into Eq.~(\ref{eq:lik}) and taking the negative logarithm yields the negative log-likelihood to be minimized.
\begin{equation}
-\ln\mathcal{L}(x;\boldsymbol{\theta}) = -\sum_{i=1}^{N_{\rm data}} \ln I_i(\boldsymbol{\theta})
\;+\; N_{\rm data}\ln\sigma(\boldsymbol{\theta}),
\label{eq:nll}
\end{equation}
where the first term evaluates the intensity at the observed events, and the second term normalizes the model to the phase-space integral through the Monte Carlo estimate in Eq.~(\ref{eq:sigma}). The minimum of Eq.~(\ref{eq:nll}) determines $\boldsymbol{\theta}$, namely the couplings and resonance parameters, from the measured four-momenta.
The event intensity is defined as the squared total amplitude, summed over all physical polarizations.
\begin{equation}
I(x;\boldsymbol{\theta}) = \sum_{\boldsymbol{\lambda}}
\left|\sum_{a} \boldsymbol{\theta}_{a} \mathcal{A}_{\boldsymbol{\lambda},a}(x;\boldsymbol{\theta}_r)\right|^2,
\label{eq:amp-total}
\end{equation}

In the covariant orbital-spin ($L$--$S$) tensor formalism~\cite{Zou:2002ar,Jing:2023rwz,Jing:2024mag,Li:2022qff}, the two-body vertices in a decay chain are tensors constructed from the four-momenta and polarization tensors of the particles, without requiring chains of Lorentz boosts and rotations to connect the helicity frames of successive decays. For a chain $a$ characterized by quantum numbers $\{L_i,S_i\}$ at each vertex, the amplitude for an event with polarization configuration $\boldsymbol{\lambda} \equiv \{\sigma_1,\sigma_2,\ldots\}$ factorizes into the chain coupling, the propagator $F(m;\boldsymbol{\theta}_r)$ of the resonance $r$ at the invariant mass $m$ of its daughter system, and a kinematic factor that carries the spin structure of the chain,
\begin{equation}
\mathcal{A}_{\boldsymbol{\lambda},a}(x;\boldsymbol{\theta}_r)
= \, F(m;\boldsymbol{\theta}_r) \, \sum_{\sigma_X} [\mathcal{A}_{L_1,S_1}]_{\sigma_1}^{\sigma_2\sigma_X} [\mathcal{A}_{L_2,S_2}]_{\sigma_X}^{\sigma_3\sigma_4}\cdots,
\label{eq:amp-factor}
\end{equation}
The kinematic factor, where $[\mathcal{A}_{L,S}]$ denotes the vertex tensors of the chain, depends only on the event four-momenta $x$ and on the quantum numbers fixed by the decay topology, and never on the fit parameters $\boldsymbol{\theta}$. These parameters enter the chain amplitude only through $\boldsymbol{\theta}_{a}$ and $F(m;\boldsymbol{\theta}_r)$. Therefore, all kinematic factors can be evaluated once before the fit and reused at every iteration. The computational cost of a fit is dominated by evaluating $I(x;\boldsymbol{\theta})$ for every event at every iteration. The next section describes how \texttt{CTPWA} organizes this evaluation.

\section{Computational optimization}
\label{method}

\texttt{CTPWA} implements several optimizations to achieve high-performance partial-wave analysis. To overcome the memory and CPU--GPU communication challenges of conventional frameworks, this section describes two key strategies. First, all parameter-independent kinematic quantities are precomputed and cached once, so device memory usage is fixed before the fit starts, and kinematic factors are not recomputed. Second, the entire likelihood, its analytic derivatives, and the minimization loop are implemented entirely on the GPU, eliminating host--device transfers, kernel-launch and graph-tracing overheads, and per-iteration allocations. As a result, each iteration evaluates only the small parameter-dependent part of the model, and device memory usage remains constant throughout the fit.

\subsection{Precomputation and caching mechanisms}

The storage design of \texttt{CTPWA} follows directly from the factorization in Eq.~(\ref{eq:amp-factor}): all quantities entering the fit are divided into two classes with different lifetimes.

The first class comprises the parameter-independent vertex tensors $\mathcal{A}_{L,S}$, which depend only on the event kinematics and the decay topology. They are evaluated once at initialization from the event four-momenta, stored in contiguous arrays in device memory, and kept resident for the entire fit. Their memory usage is fixed and known before the fit starts. They are used in every iteration and derivative evaluation, and no quantity in this class is ever recomputed or reallocated.

The second class comprises the fit parameters themselves. They are stored as a single small vector of $N_{\boldsymbol{\theta}}$ numbers and updated once per iteration; this vector is the only quantity that changes during minimization.

The two classes interact minimally. When the resonance parameters are fixed, the propagators are constants, and therefore the channel amplitudes are fixed as well: the entire amplitude model for each event reduces to a small matrix computed once, and each iteration requires only updating the coupling vector and multiplying it by the stored matrices. When the resonance parameters float, the vertex tensors are still precomputed once; each iteration recomputes only the scalar propagators $F(m;\boldsymbol{\theta}_r)$, one complex number per event per resonance, and multiplies them by the precomputed tensors, with the couplings applied last: coupling $\times$ precomputed $\mathcal{A}_{L,S}$ $\times$ $F$.

This strict separation is the core of the memory optimization. The dominant memory content is written once at initialization and subsequently only read, and the per-iteration work accesses only the small parameter vector and the current propagators. No allocation occurs during the fit, no kinematic factor is recomputed, and no event data are transferred between the host and the device.

\subsection{Fully GPU-based likelihood computation and minimization}

The events in a PWA sample are mutually independent. According to Eq.~(\ref{eq:nll}), the negative log-likelihood is a sum of per-event contributions, and the intensity of each event can be evaluated independently, making the likelihood computation embarrassingly parallel. GPUs are a natural platform for this task, but event-by-event amplitude evaluation underutilizes their computational capacity. The arithmetic workload per event is too small to saturate the device, and the computation is dominated by memory latency and branching over decay chains and spins. In contrast, matrix operations are the native workload of GPUs and are performed by highly optimized linear algebra kernels. The design principle of \texttt{CTPWA} is therefore to express the likelihood and all its derivatives as batched dense linear algebra over the event dimension, such that each event is an independent row and every computation maps to optimized matrix kernels without per-event branching.

The channel amplitudes for each polarization configuration form a complex vector $\boldsymbol{\mathcal{A}}_{\boldsymbol{\lambda}}$ of length $N_{\rm ch}$, and the event intensity becomes a quadratic form in the coupling parameters $\boldsymbol{\theta}_a$,
\begin{equation}
I(x;\boldsymbol{\theta}) = \sum_{\boldsymbol{\lambda}}
\bigl|\boldsymbol{\theta}_a\boldsymbol{\mathcal{A}}_{\boldsymbol{\lambda}} \bigr|^2
= \boldsymbol{\theta}_a^\dagger \mathbf{B} \boldsymbol{\theta}_a, \qquad
\mathbf{B} \equiv \sum_{\boldsymbol{\lambda}}
\boldsymbol{\mathcal{A}}_{\boldsymbol{\lambda}}^* \boldsymbol{\mathcal{A}}_{\boldsymbol{\lambda}}^{\mathsf{T}},
\label{eq:quadratic}
\end{equation}
where $\mathbf{B}$ is an $N_{\rm ch}\times N_{\rm ch}$ Hermitian positive-semidefinite matrix for each event. All kinematic and polarization information for an event is incorporated into $\mathbf{B}$. Thus, in each iteration, the intensity update for the entire data set reduces to a batched matrix--vector product, and the negative log-likelihood in Eq.~(\ref{eq:nll}) is obtained from the elementwise logarithms of the resulting intensities. For phase-space normalization, the per-event matrices are averaged once in advance,
\begin{equation}
\bar{\mathbf{B}} = \frac{1}{N_{\rm phsp}} \sum_{j=1}^{N_{\rm phsp}} \mathbf{B}_j,
\label{eq:Bbar}
\end{equation}
so that the phase-space contribution to the likelihood and its derivatives require only a single $\mathcal{O}(N_{\rm ch}^2)$ operation per iteration, independent of $N_{\rm phsp}$. 

All derivatives are evaluated from the same stored matrices in a batched pass over events. For derivatives with respect to the coupling parameters $\boldsymbol{\theta}_a$, the data contribution to the gradient of Eq.~(\ref{eq:nll}) is
\begin{equation}
\frac{\partial\mathcal{L}_{\rm data}}{\partial\boldsymbol{\theta}_a}
= -\frac{2\mathbf{B}\boldsymbol{\theta}_a}{I},
\label{eq:grad-matrix}
\end{equation}
The division is performed elementwise across events. The phase-space contribution is obtained analogously through $\bar{\mathbf{B}}$.

When the resonance parameters $\boldsymbol{\theta}_r$ are allowed to vary, the derivative of the negative log-likelihood with respect to a specific parameter $\theta_{r}$ of resonance $r$ takes the form
\begin{equation}
\frac{\partial\mathcal{L}_{\rm data}}{\partial\theta_{r}}
= -2 \sum_{\boldsymbol{\lambda}} \frac{1}{I_{\boldsymbol{\lambda}}}
\mathrm{Re}\left[
\mathcal{P}_{r,\boldsymbol{\lambda}} \, \bar{S}_{\boldsymbol{\lambda}} \,
\frac{\partial F(m;\boldsymbol{\theta}_r)}{\partial\theta_{r}}
\right],
\label{eq:grad-r}
\end{equation}
where $\mathcal{P}_{r,\boldsymbol{\lambda}} \equiv \sum_{a\in r} \theta_{a} [\mathcal{A}_{L,S}]_{a,\boldsymbol{\lambda}}$, with $[\mathcal{A}_{L,S}]_{a,\boldsymbol{\lambda}}$ denoting the precomputed kinematic component of chain $a$, and $S_{\boldsymbol{\lambda}}=\sum_a \theta_{a}\mathcal{A}_{\boldsymbol{\lambda},a}$ denoting the total amplitude for the polarization configuration $\boldsymbol{\lambda}$.
The phase-space normalization contributes an analogous term through the averaged matrices $\bar{\mathbf{B}}$; its closed-form expression is obtained by replacing $\sum_{\boldsymbol{\lambda}}$ with the phase-space average and differentiating Eq.~(\ref{eq:sigma}) accordingly.
In practice, the inner product $\mathcal{P}_{r,\boldsymbol{\lambda}}$ is independent of the specific parameter $\theta_{r}$ and is computed once per event per resonance. Only the derivative of the scalar propagator, $\partial F/\partial\theta_{r}$, is evaluated anew using lightweight automatic forward-mode differentiation. This structure allows the full gradient with respect to all resonance parameters to be accumulated in a single batched pass over events, without any per-parameter backward traversal.

The per-event contribution to the real Hessian naturally separates into three blocks according to parameter type: coupling--coupling ($aa$), coupling--resonance ($ar$), and resonance--resonance ($rr$). 
The coupling--coupling block has the compact analytic form
\begin{equation}
\mathbf{H}_{aa} = -\frac{2\mathbf{B}}{I}
+ \frac{4}{I^2} \, (\mathbf{B}\boldsymbol{\theta}_a)(\mathbf{B}\boldsymbol{\theta}_a)^{\mathsf{T}},
\label{eq:hess-cc}
\end{equation}
which is a scaled copy of $\mathbf{B}$ that has already been formed during the intensity evaluation.
The coupling--resonance block is given by
\begin{equation}
\mathbf{H}_{ar} = -\frac{2}{I}\,\mathrm{Re}\left( \bar{S} \cdot \nabla_{\boldsymbol{\theta}_r} \mathbf{d}S_a^\mathsf{T} \right)
+ \frac{4}{I^2} \left[ \mathrm{Re}(\mathbf{B}\boldsymbol{\theta}_a) \right] \left[ \mathrm{Re}\left( \bar{S} \cdot \nabla_{\boldsymbol{\theta}_r} \mathbf{d}S_a^\mathsf{T} \right) \right]^\mathsf{T},
\label{eq:hess-cr}
\end{equation}
where $\mathbf{d}S_a = \partial S/\partial\boldsymbol{\theta}_a$. This block involves only first-order derivatives of the amplitude.
For the resonance--resonance block
\begin{equation}
\mathbf{H}_{rr} = -\frac{2}{I}\,\mathrm{Re}\left( \bar{S} \cdot \nabla_{\boldsymbol{\theta}_r}\nabla_{\boldsymbol{\theta}_r}^\mathsf{T} S \right)
+ \frac{4}{I^2} \left[ \mathrm{Re}\left( \bar{S} \cdot \nabla_{\boldsymbol{\theta}_r} S \right) \right]
\left[ \mathrm{Re}\left( \bar{S} \cdot \nabla_{\boldsymbol{\theta}_r} S \right) \right]^\mathsf{T}.
\label{eq:hess-rr}
\end{equation}
The second-derivative term $\nabla_{\boldsymbol{\theta}_r}\nabla_{\boldsymbol{\theta}_r}^\mathsf{T} S$ is block-diagonal across resonances because cross terms between different resonances vanish. Thus, only the diagonal sub-blocks corresponding to individual resonances receive second-derivative contributions.

The full Hessian is accumulated in a single batched pass over events. This analytic construction avoids the $\mathcal{O}(N_{\rm params})$ backward passes required by automatic differentiation and makes Hessian-based uncertainty estimation practical even for high-statistics data sets.

In conventional pipelines, minimization is performed on the CPU, while likelihood evaluation is offloaded to the GPU. At each iteration, the events or recomputed amplitudes are transferred to the device, and the resulting likelihood value is copied back to the host. Consequently, data and intermediate results repeatedly cross the host--device boundary, and transfer time often dominates the iteration cost. In contrast, \texttt{CTPWA} performs the entire optimization on the GPU. The negative log-likelihood, its analytic gradient, and its Hessian are exposed directly to the L-BFGS optimizer~\cite{Liu:1989esw} in PyTorch~\cite{pytorch}, which runs on the device. All precomputed tensors and resident data quantities remain in GPU memory throughout the fit. No computational graph is built during optimization, because the computationally intensive derivatives are analytic and only the scalar propagator functions use lightweight forward-mode automatic differentiation. The analytic gradient has been cross-checked against PyTorch autograd on the same likelihood and agrees to machine precision.

\section{Performance and demonstration}
\label{bench}

\subsection{Computational performance}

To quantify the performance of the proposed method, we benchmark it using synthetic PWA samples with $N_{\rm ch}=10$ decay channels, $N_{\rm pol}=3$ polarization configurations, and $N_{\rm phsp}=10\times N_{\rm data}$ phase-space events, in double precision. Three approaches are compared: the full amplitude calculation on the CPU within PyTorch~\cite{pytorch}, the same PyTorch-based computation on the GPU, and the optimized formulation, in which the kinematic amplitudes are precomputed once and the gradient and Hessian are obtained analytically. This comparison separates the impact of hardware acceleration from the algorithmic improvements. All timings are averaged over repeated evaluations after warm-up.

\begin{figure}[H]
\centering
\includegraphics[width=0.4\textwidth,page=3]{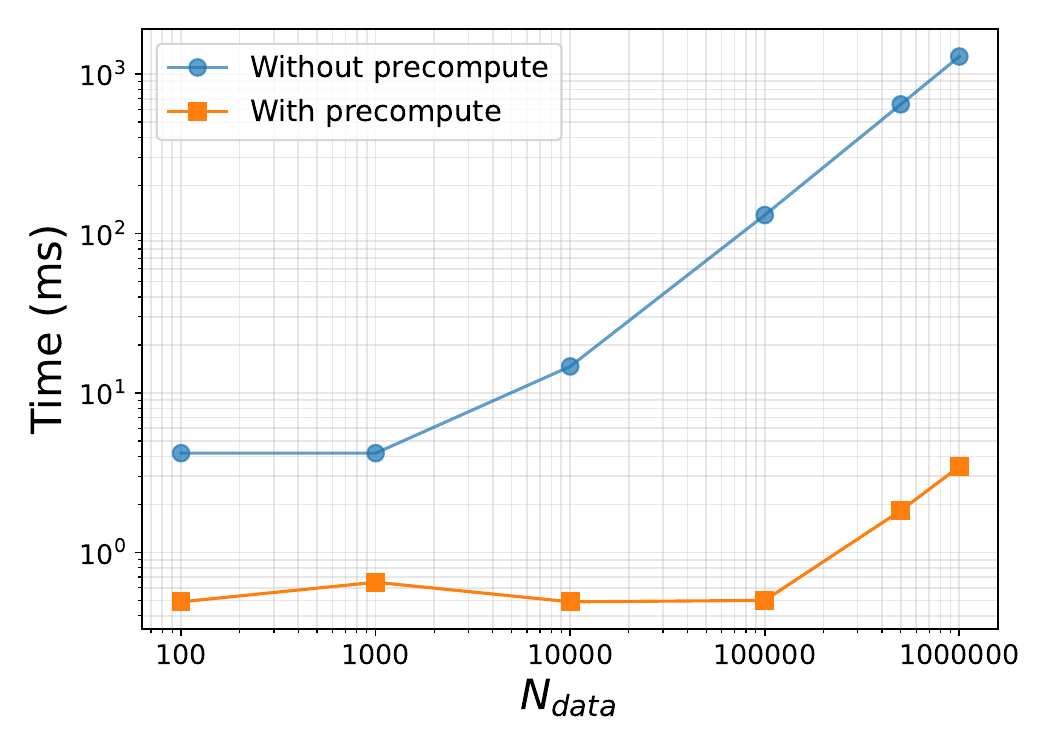}
\caption{Per-iteration time as a function of $N_{\rm data}$, comparing recomputing amplitudes at each iteration with the precomputed method. The runtime of the precomputed method is nearly independent of the data size, whereas that of the conventional pipeline increases linearly and eventually exhausts the GPU memory.}
\label{fig:pre-scale}
\end{figure}

Figure~\ref{fig:pre-scale} compares the per-iteration time of the conventional pipeline with that of the precomputed method for a typical 4-body decay topology. The per-iteration cost of the conventional approach increases linearly with the data size, from about 4\,ms at $10^2$ events to about 1.2\,s at $10^6$ events. By contrast, the precomputed pipeline remains essentially constant at about 0.5--3.5\,ms, because only batched matrix--vector operations remain. Including the one-time precomputation, a complete fit of $10^3$ iterations is accelerated by a factor that increases from about 8 at $N_{\rm data}=10^2$ to more than two orders of magnitude at $N_{\rm data}=10^6$. The memory usage shows a similarly pronounced advantage. Beyond $N_{\rm data}=10^6$, the conventional pipeline exhausts the device memory, whereas the precomputed representation completes without difficulty.

\begin{figure}[H]
\centering
\includegraphics[width=0.4\textwidth,page=1]{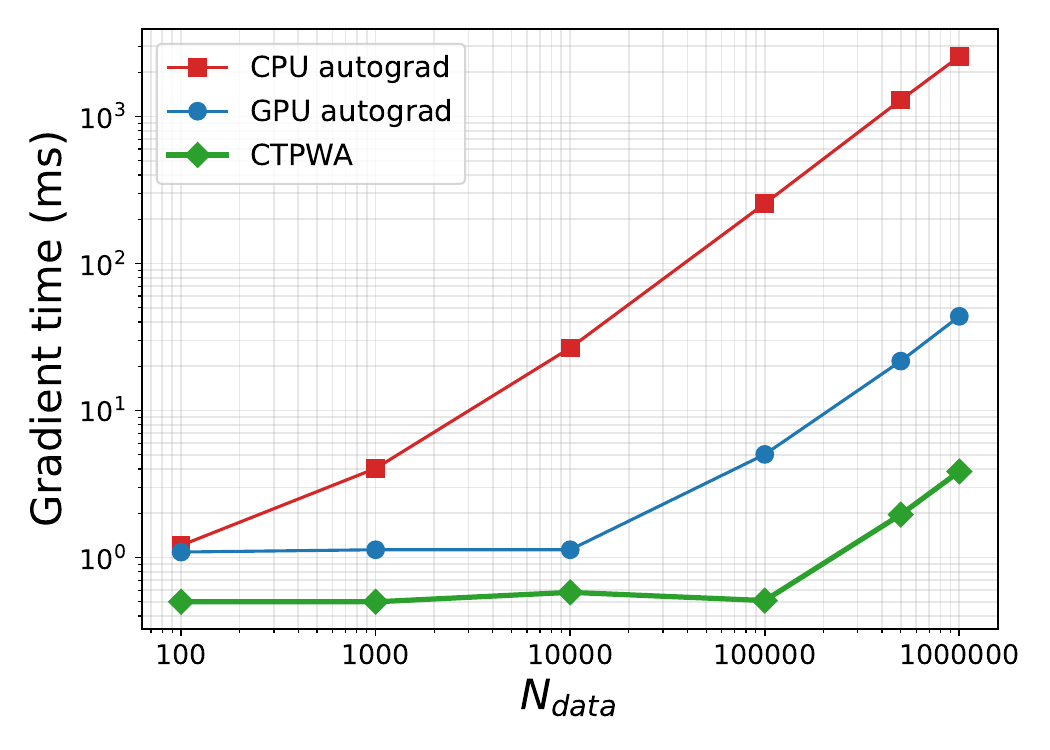}
\caption{NLL$+$gradient evaluation time as a function of $N_{\rm data}$. The red, blue, and green curves correspond to CPU-based PyTorch autograd, GPU-based PyTorch autograd, and the analytic formulation implemented in \texttt{CTPWA}, respectively. The analytic formulation represented by the green curve is consistently faster by one to two orders of magnitude.}
\label{fig:fix-grad}
\end{figure}

Because the matrix formulation in Sec.~\ref{method} evaluates the likelihood and its gradient using batched matrix operations, the optimized formulas compute the NLL and gradient for $10^6$ events in approximately 4\,ms, compared with 45\,ms for GPU autograd and 2.5\,s for CPU autograd, as shown in Fig.~\ref{fig:fix-grad}. The speedup increases with data size: at $10^5$ events, it is about 9-fold relative to GPU autograd and about 500-fold relative to CPU autograd, increasing to about 11-fold and about 660-fold, respectively, at $10^6$ events. This favourable scaling reflects the weak dependence of the batched matrix--vector operations on $N_{\rm data}$, in contrast to the per-event overhead of graph-based differentiation.

\begin{figure}[H]
\centering
\includegraphics[width=0.4\textwidth,page=2]{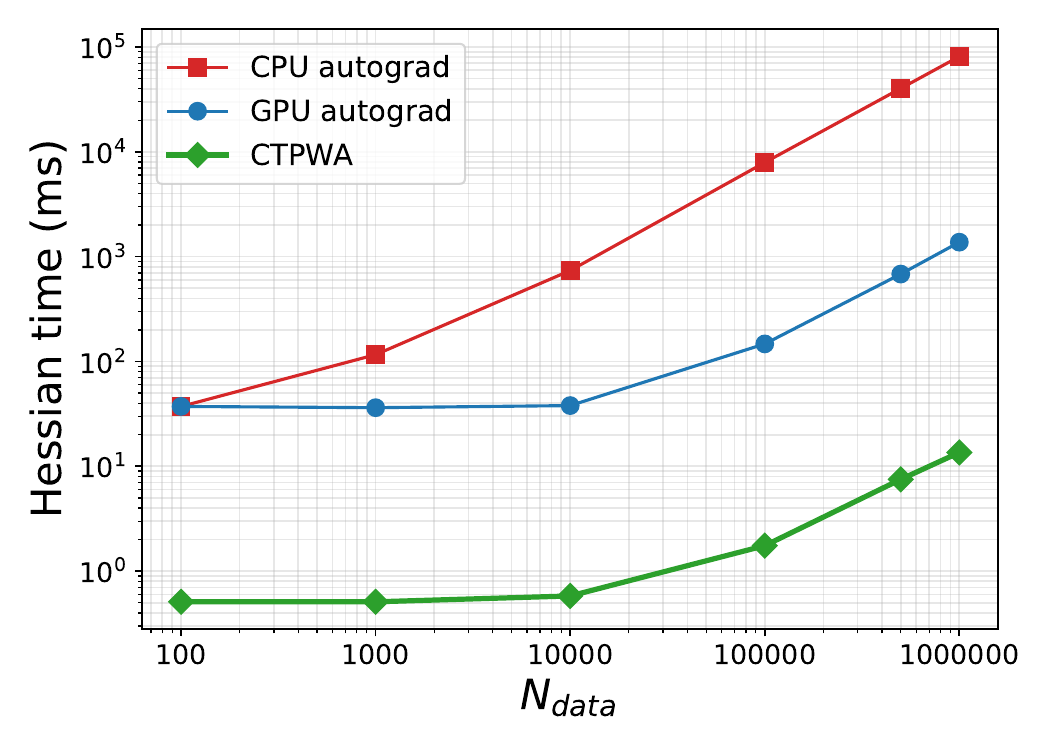}
\caption{Hessian evaluation time as a function of $N_{\rm data}$. The red, blue, and green curves correspond to CPU-based PyTorch autograd, GPU-based PyTorch autograd, and the analytic formulation of \texttt{CTPWA}, respectively. The analytic method is faster by two to three orders of magnitude across the full range.}
\label{fig:fix-hess}
\end{figure}

The greatest performance gain occurs in the Hessian evaluation, as shown in Fig.~\ref{fig:fix-hess}. For $10^6$ events, the analytic Hessian of Sec.~\ref{method} requires approximately 13\,ms for the full $20\times20$ matrix, compared with 1.4\,s for GPU autograd and 81\,s for CPU autograd, corresponding to speedup factors of approximately 101 and $6\times10^3$, respectively. The single-pass accumulation replaces the $\mathcal{O}(N_{\rm params})$ backward passes required by automatic differentiation, making Hessian-based uncertainty estimation practical at high statistics.

\subsection{Fit demonstration with $J/\psi\to K^+K^-\eta$}
\label{kketa}

As a demonstration of the framework in a realistic setting, we perform a fit to the decay $J/\psi\to K^+K^-\eta$ with a sample of one million toy Monte Carlo events. This three-body decay probes intermediate resonances in both the $K^+K^-$ and $K\eta$ subsystems, making it a representative test case for modern PWA studies. The amplitude model includes 16 intermediate resonances selected from the PDG~\cite{ParticleDataGroup:2024cfk} with 33 complex couplings, and the phase-space integral is evaluated with $5\times10^6$ uniformly generated Monte Carlo events. The kinematic amplitudes are precomputed once in about 94\,s and stored in GPU memory. We consider two scenarios: (i) resonance masses and widths fixed to their input values, and (ii) ten resonance parameters allowed to float together with the couplings.

Table~\ref{tab:demo-timing} reports the measured wall-clock time for both scenarios. In the case with fixed resonance parameters, the per-iteration cost is dominated by the batched matrix--vector products in Eqs.~(\ref{eq:quadratic}) and (\ref{eq:grad-matrix}), and the full fit is completed in about 7\,s after about 500 iterations. When resonance parameters are floated, the phase-space pre-averaging in Eq.~(\ref{eq:Bbar}) no longer applies, and the gradient must be recomputed for each event, increasing the per-iteration cost to about 100\,ms. The fit still converges in about 100\,s after about 1000 iterations. The Hessian evaluation time increases from about 0.35\,s to about 1.6\,s due to the additional resonance-parameter blocks.

\begin{table}[H]
\centering
\caption{
Measured wall-clock time and device memory usage for the $J/\psi\to K^+K^-\eta$ fit with $N_{\rm data}=10^6$ and $N_{\rm phsp}=5\times 10^6$. The two columns correspond to the cases in which the resonance parameters are held fixed and allowed to float, respectively.
}
\label{tab:demo-timing}
\begin{tabular}{lll}
\toprule
Stage & Fixed resonance params & Float resonance params \\
\midrule
Amplitude initialization & $\sim$94\,s & $\sim$94\,s \\
NLL $+$ gradient per iteration     & $\sim$15\,ms & $\sim$100\,ms \\
Full Hessian            & $\sim$0.35\,s & $\sim$1.6\,s \\
Device memory           & $\sim$20\,GB & $\sim$32\,GB \\
Complete single fit     & $\sim$7.0\,s & $\sim$100\,s \\
\bottomrule
\end{tabular}
\end{table}

\begin{figure}[H]
\centering
\includegraphics[width=0.7\textwidth,page=1]{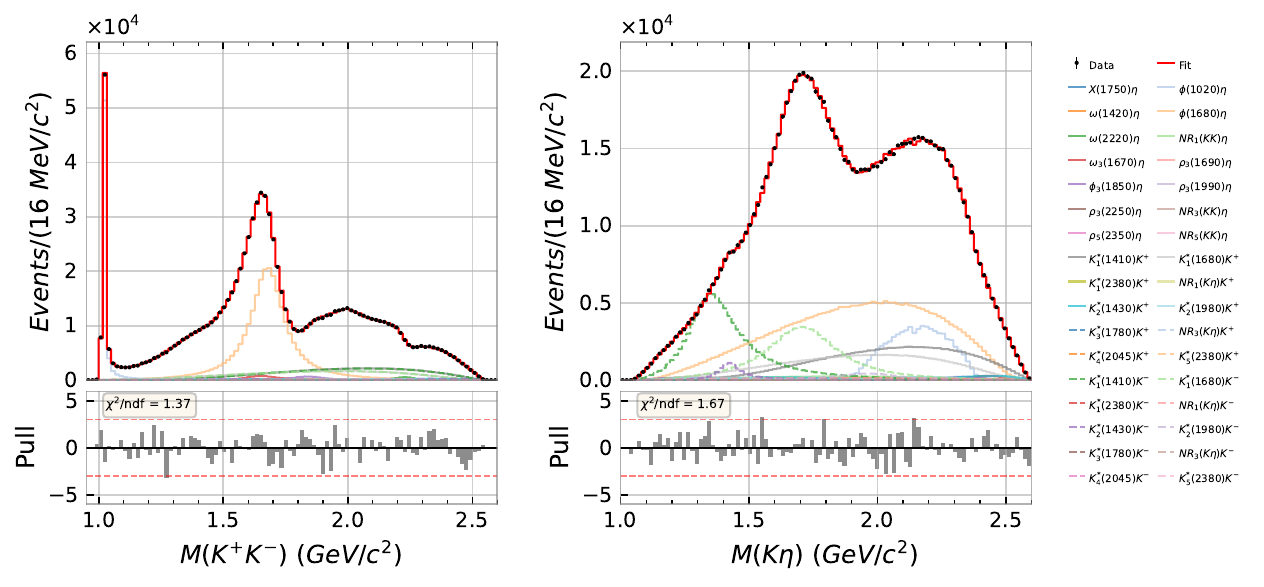}
\caption{Fit results for $J/\psi\to K^+K^-\eta$. The data points represent the signal Monte Carlo sample, and the solid histogram shows the projection of the best-fit model. The left and right panels show $M(K^+K^-)$ and $M(K\eta)$, respectively.}
\label{fig:demo-mass}
\end{figure}

Figure~\ref{fig:demo-mass} shows the invariant-mass projections of the best-fit model superimposed on the signal Monte Carlo sample. The model accurately reproduces the resonant structures across the full kinematic range. The residuals are consistent with statistical fluctuations, and the fitted parameters agree with the generated input values. A complete high-statistics amplitude fit can thus be completed within minutes, a performance level that readily meets the typical requirements of BESIII partial-wave analyses.

\section{Summary}
\label{summary}
We have developed \texttt{CTPWA}, a GPU-oriented partial wave analysis framework for high-statistics hadron spectroscopy. It addresses the main limitations of conventional unbinned maximum-likelihood fits, in which repeated event-by-event amplitude evaluation incurs substantial computational time and memory requirements, especially for samples of $10^5$--$10^6$ events and models with many interfering amplitudes.

The method exploits the factorization of the covariant tensor amplitude formalism. The kinematic tensors depend only on the event four-momenta and decay topology and are therefore independent of the fit parameters. They are precomputed once, stored in device memory, and reused throughout the fit, while only the couplings and, when allowed to float, the resonance propagators are updated during minimization. The likelihood is rewritten in matrix form, enabling the negative log-likelihood, gradient, and Hessian to be evaluated using batched dense linear algebra on the GPU. The dominant derivatives are obtained analytically, avoiding the overhead of graph-based automatic differentiation, and the full minimization loop is executed on the device.

The framework demonstrates substantial gains in both speed and memory efficiency. For synthetic samples with $10^6$ events, the optimized implementation evaluates the negative log-likelihood and gradient in a few milliseconds, approximately one order of magnitude faster than GPU autograd and more than three orders of magnitude faster than CPU autograd. The analytic Hessian is approximately two orders of magnitude faster than GPU autograd and more than four orders of magnitude faster than CPU autograd. The precomputed representation also avoids the memory exhaustion encountered in conventional iterative amplitude recomputation. To further validate the framework in a realistic application, we fit $J/\psi\to K^+K^-\eta$ using Monte Carlo samples of $10^6$ signal events and $5\times10^6$ phase-space events. A fit with fixed resonance parameters converges within seconds, while a fit with floating resonance parameters completes in about $10^2$ seconds. The fitted projections reproduce the generated distributions well, and the extracted parameters are consistent with the input values.

These results demonstrate that \texttt{CTPWA} provides an efficient solution to the computational challenges of large-scale PWA at modern experiments such as BESIII and LHCb. In addition, this framework is expected to improve substantially the efficiency of resonance searches, precision measurements, and systematic model comparisons and to provide a practical basis for AI-assisted studies in hadron spectroscopy in the near future.

\section{Acknowledgments}
We thank Prof. Haojie Jing and Prof. Jiajun Wu for useful discussions.
This work is supported in part by the National Key R\&D Program of China under Contract No. 2025YFA1613900; the National Natural Science Foundation of China (NSFC) under Contract Nos. 12225509 and 12235017; and the Strategic Priority Research Program of the Chinese Academy of Sciences under Grant No. XDA0480600.

\section{Program availability}

The \texttt{CTPWA} framework described in this paper is publicly available at \url{https://github.com/BHXiang/ctpwa}.

\vspace{-2.5mm} \centerline{\rule{80mm}{0.1pt}} \vspace{1mm}

\bibliographystyle{apsrev4-2}
\bibliography{main}

\end{document}